\documentclass[12pt]{iopart}
\input epsf
\usepackage{epsf,graphicx,iopams}  
\def\msun{$M_{\odot}$}

\def\lsim{\mathrel{\rlap{\lower4pt\hbox{\hskip1pt$\sim$}}
    \raise1pt\hbox{$<$}}}

\begin{document}

\title[Coherent Network Detection of Gravitational Waves: The Redundancy Veto]{Coherent Network Detection of Gravitational Waves: The Redundancy Veto}

\author{Linqing Wen and Bernard F Schutz}
\address{Max Planck Institut fuer Gravitationsphysik, Albert-Einstein-Institut,
Am Muehlenberg 1,  D-14476 Golm, Germany}
\begin{center}
\begin{abstract}
A network of gravitational wave detectors is called {\em redundant} if, given 
the direction to a source, the strain induced by a gravitational wave in 
one or more of the detectors can 
be fully expressed in terms of the strain induced in others in the network. Because 
gravitational waves have only two polarizations, any network of three or more
differently oriented interferometers with similar observing bands 
is redundant. The three-armed LISA space 
interferometer has three outputs that are redundant at low frequencies. The 
two aligned LIGO interferometers at Hanford WA are redundant, and 
the LIGO detector at Livingston LA is nearly redundant with either of the Hanford 
detectors. Redundant networks have a powerful veto against spurious noise, 
a linear combination of the detector outputs that contains {\em no}
gravitational wave signal. For LISA, this ``null'' output is known as the Sagnac
mode, and its use in discriminating between detector noise and a cosmological 
gravitational wave background is well understood \cite{tinto1,tinto2}. But the usefulness of the 
null veto for ground-based detector networks has been ignored until now. We 
show that it should make it possible to discriminate in a model-independent way 
between real gravitational waves and accidentally coincident non-Gaussian 
noise ``events'' in redundant networks of two or 
more broadband detectors.  It has been shown that with three detectors, the null 
output can even be used to locate the direction to the source, and then two other 
linear combinations of detector outputs give the optimal ``coherent'' 
reconstruction of the two polarization components of the signal. We discuss 
briefly the implementation of such a detection strategy in realistic networks, where 
signals are weak, detector calibration is a significant uncertainty, and 
the various detectors may have different (but overlapping) observing bands.
\end{abstract}
\end{center}



\maketitle

\section{Introduction}
\label{intro}
We are entering a new phase in the development of 
gravitational wave astronomy. The 
new generation of large-scale gravitational wave (GW) 
interferometers has begun operating.  In particular, the American LIGO 
detectors \cite{LIGO_GWDAW} are very close to their first sensitivity 
goals, having surpassed the older generation of cryogenic bar 
detectors \cite{fafone04} 
in sensitivity to almost all potential sources in a 
broad intermediate frequency band, from a 
few tens of Hz to several kilohertz. LIGO and its German-British partner 
GEO600 \cite{GEO_GWDAW} have taken data so far in four ``science runs'', 
sometimes also in cooperation with the Japanese TAMA 
detector \cite{TAMA_GWDAW}. (For upper limits from the earliest 
of these runs, see \cite{S1_bursts,S1_pulsars,S1_inspiral,S1_stoch,S2_pulsars}.) 
The Italian-French VIRGO 
detector \cite{VIRGO_GWDAW} is currently undergoing commissioning. By the 
end of 2005, LIGO and GEO are expected to have embarked on full-time 
observing. Within a year or two we should know whether or not the 
first-stage sensitivity of these detectors is sufficient to make the first 
detections of gravitational waves, or whether the field will have to wait
for the sensitivity upgrades that are planned over the subsequent five years.

The principal problem facing GW detection is distinguishing weak signals
from detector noise. The main sources of noise in detectors ideally produce 
Gaussian (normally-distributed) amplitude noise, but all detectors 
have poorly understood 
sources of ``non-Gaussian noise'', characterized by large-amplitude 
disturbances that occur much more frequently 
than in a Gaussian distribution. These disturbances are unpredictable 
and it is often difficult, if not impossible, to trace them to their 
cause, which may be external to the detector or the result of a temporary 
malfunction within the detector. Genuine GW signals, of course,
 are also expected to be rare, so they can in principle be 
confused with non-Gaussian 
noise events. Even if the waveform of an expected signal 
is known ahead of time, 
there is some chance that noise will match it well enough to confuse.

The optimal way to detect signals with a network of detectors that have ideal 
Gaussian noise backgrounds is called ``coherent detection'', and has been studied
by a number of authors \cite{krolak94, bose99,pai01,finn01}. If the direction to 
a source is known, then one forms linear combinations of the detector outputs, 
with suitable time-delays, that best give the amplitudes of the two independent
gravitational-wave polarizations. These superposed data streams can be studied 
for signals using matched filtering or less specific methods, like wavelet 
transforms or time-frequency methods. Gravitational waves are identified if 
these superposed streams contain excursions that would have very low probability 
in the purely Gaussian noise of the detectors.

Unfortunately, this coherent method is vulnerable to confusion through a 
non-Gaussian noise event in a single detector, which could make an unexpectedly 
large excursion in the superpositions that represent the GW 
polarizations. Therefore, current searches use coincidence testing, where 
individual data streams must each pass an amplitude threshold (usually after 
processing the data stream in some way); only coincident events are taken to be  
candidates for GWs \cite{S1_bursts,LIGO_TAMA_GWDAW}. This cuts down the confusion from 
non-Gaussian noise, since these relatively rare noise events do not often occur 
at the same time (or with suitable time-delays for the wave travel time) in 
different detectors. But the coincidence method is clearly non-optimal for a general network of detectors, since the quadrupole 
antenna-patterns of differently oriented
detectors will respond differently to a real GW, so that in some 
cases the detector outputs will not in fact all cross the pre-defined amplitude 
threshold in coincidence, and the wave will not be recognized (a {\em false 
dismissal}\;).  What is more, 
a threshold criterion on its own is a crude identifier for a GW: 
it does not use all the information about the waveform that is present in 
the data streams, and so it can lead to {\em false detections}.
More sophisticated analysis methods based on phase correlation of triggered events from
different detectors can be found in \cite{laura, sergey}

We discuss here a simple analysis method designed to protect the coherent 
method from confusion by spurious noise. First introduced by G\"ursel 
\& Tinto \cite{Guersel} in 1989, it works in a network of detectors 
that contains enough information to construct what we call a 
{\em null stream},  by which we mean  
a particular linear combination of all the available data 
that {\em cancels out} a 
GW signal from a particular sky direction. We call such 
networks {\em redundant}, since they contain enough information to reconstruct 
the response of one detector from the responses of the others, regardless
of the polarization of the wave.  The null stream in a 
redundant network is a consistency check to test whether a candidate GW  event 
is produced by detector noise or by a real GW. If the null stream shows 
an unusual excitation, then the candidate should be rejected (vetoed), 
since only 
noise can excite the null stream. Conversely, if individual detectors 
register an excitation that is absent from the null stream, then it is very 
likely to be a real gravitational wave, since cancellation in the null stream 
of a noise-generated event would require a highly improbably coincidence 
in amplitude and phase among all the detectors in the network. In practice, 
all current and planned networks of detectors are redundant or nearly so, 
which means that the null-stream test can be implemented as a veto in upcoming 
realistic searches.

Crucially, the null stream provides a {\em model-independent} veto, 
depending only on the description 
of GWs provided by general relativity. (We will discuss briefly the 
way in which the null stream can actually be used to {\em test} the GW 
model of general relativity.) The null stream does not require 
a source model, an assumption about polarization, 
or a transfer function that shows how a particular disturbance 
can affect the data stream of a detector. If the network is redundant, then 
the null stream constructed for a particular source direction cannot contain 
any signal from sources in that direction. This veto can be used to supplement 
thresholding in a coincidence search.  This method is different from
cross-correlation methods in that it checks the consistency in both the
amplitude and phase. We expect that it can  
dramatically reduce the number of false identifications.

G\"ursel \& Tinto argue that the null stream can be used to identify 
the locations of sources on the sky (and therefore the polarizations
of GWs) in the case of 3-detectors at different sites, since it will consist of pure noise only
when constructed for the correct source location. We have performed simulations
that bear this out. 

In practice, the null stream veto has limitations. For example, detectors
in a redundant network may have non-overlapping bandwidths, or one of 
the detectors may be significantly less sensitive than the others. In these
cases the null stream is self-limiting: it will contain so much noise 
that a GW will not be visible, making the stream useless. For this reason 
the null stream is unlikely to be useful for networks of bar detectors, or 
in networks where there are big differences in detector sensitivity. 
Moreover, the null stream is sensitive to calibration errors, and calibrating 
gravitational wave detectors is difficult. Therefore residuals of real 
signals may remain in the null stream at the ten percent level, which is 
the typical calibration uncertainty at present.

Implementing the null-stream veto essentially means rejecting events if 
the null stream contains an unexpectedly large excitation, which 
must be due to ``non-Gaussian'' 
noise events in at one or more detectors. This imposes the requirement 
that a GW can only be identified with confidence  if the event arrives at 
a moment when all 
the detectors are operating in their ``normal'' state, i.e.\ with purely 
Gaussian noise (after removal of known instrumental artifacts,
e.g.,  lines). Since non-Gaussian events are rare, and GW events are even more
rare, it seems safe to impose this requirement. Indeed, it is difficult to 
see how a first GW detection could be claimed with any confidence 
if it were known that at least one of 
the detectors itself had a simultaneous non-Gaussian noise event. 
It follows that a crucial implementation issue for 
the null-stream veto is the choice of a criterion for whether 
the null stream is consistent with Gaussian noise at the time of an event. 
In this paper we use a chi-squared test for our simulations, but the 
actual test may depend on the kind of processing of the individual 
detector streams that took place before the ``event'' was recognized.

In this paper, we explore the null-stream veto and detector for 
unknown waveforms (i.e., the so-called burst GWs) observed by interferometers, 
although the veto could be applied to any type of signal.  
We consider three realistic cases: the two LIGO Hanford 
detectors (the four-km H1 and the two-km H2), which form a redundant network 
of just two detectors; a three-detector redundant 
network consisting of H1, the LIGO 
Livingston detector (L1), and GEO600; and 
an approximately redundant two-detector network consisting of H1 and L1. In
section~\ref{principle}, we discuss the principle of constructing null
stream for a network of two and three detectors.  In section~\ref{tech}, we
propose a method for the consistency check of the
``absence'' of the GW signals in the null-stream. In
section~\ref{example}, we show a few examples using
simulated data to demonstrate the performance of our approach.
In section~\ref{discussion}, we discuss the possible applications of
our method  and future work.

\section{Principle of null data stream construction}
\label{principle}
The strain created in the $i^{\rm{th}}$ interferometric detector of 
a network by a GW arriving 
from a sky direction given by the right ascension and declination 
angles $\alpha$ and $\delta$ is a linear combination of the two 
polarizations of the wave,
\begin{equation}
h_i(t)=f^{+}_i(t,\alpha,\delta)h_{+}(t)+f^{\times}_i(t,\alpha,\delta)h_{\times}(t),
\label{h_org}
\end{equation}
where $t$ is the time at the detector, $f^{+}_i(t,\alpha,\delta)$ and $f^{\times}_i(t,\alpha,\delta)$ 
are the detector's 
antenna beam pattern functions (or response) to the plus and cross
polarizations of the waves, and $h_+$ and $h_{\times}$ are the 
amplitudes of the two polarizations of the GW. The definition of the 
``$+$'' and ``$\times$'' polarizations depends on an arbitrary 
orientation angle $\psi$ defined on the sky \cite{jara98}, but does not 
depend on detector orientation. The antenna beam
patterns are explicit functions of time because the detector will 
normally change its orientation with time.  In this paper, 
we will use as a reference time the local clock time at one of 
the detectors in the network (L1 when it is included in the network). For 
other applications, such as for pulsar observations, it might 
be more robust to adopt the solar barycenter as the time-reference. 

Following the
coordinate system adopted in \cite{jara98}, we can rewrite
\begin{equation}\label{eqn:antpatt}
f^{+}_i=A_i\sin(2\psi+\xi_i),\qquad f^{\times}_i=A_i\cos(2\psi+\xi_i),
\end{equation}
where $A_i=\sqrt{f^{+}_i{}^2+f^{\times}_i{}^2}$ is the amplitude of the
antenna beam pattern, directly proportional to the observed amplitude
of the wave,  and where $\xi_i=\tan^{-1} ({f^+_i}/{f^\times_i})$ at $\psi=0$ is a
quantity that describes the detector's different response to the two
polarizations and therefore depends on detector
orientation. Eqn.~\ref{h_org} can then be re-written as,
\begin{equation}
h_i(t)=A_i\sqrt{h^2_{+}(t)+h^2_{\times}(t)}
\sin\left({2\psi+\xi_i+\xi^h (t)}\right ),
\end{equation}
where $\xi^h(t)=\tan^{-1} ({h_\times(t)}/{h_+(t)})$ is the
effective phase of the waves, and is a quantity intrinsic to the
  wave. For a circularly polarized GW of monotonic frequency $f$,
  $\xi^h(t)=2\pi f t$.

For a given incoming GW, the observed strains can therefore be
different from detector to detector in two respects.  
(1) The measured amplitude of the wave is directly
proportional to the amplitude of the detector's antenna beam pattern 
$A_i$ for the wave's direction. (2) The observed effective wave phase
(therefore the apparent wave arrival time, which is normally referred to a 
fiducial feature in the waveform) is shifted by $\xi_i$, which depends
on the detector orientation. Importantly, for an incoming wave of
dominant frequency $f_0$, the difference between the geometrical time-delay 
(wave arrival-time difference between detectors) and the measured delay 
can be as large as $1/(2f_0)$, e.g.\ 2~ms  for GW with $f_0 \sim 250$ Hz.  
This can be significant,
given that  the maximum geometrical time delays between currently operating 
detectors are between 10 and 30~ms.  Note that both quantities $A_i$ and 
$\xi_i$ are independent of the wave polarization angle $\psi$. 

In the following two subsections, we discuss how to construct the null data
stream for particular networks of two and three detectors.

\subsection{Two detectors at the same site: H1-H2}
The LIGO detectors at the Hanford site consist of two interferometers
(H1 and H2), 4 and 2 km long, sharing the same vacuum system. The 
antenna patterns of these detectors are identical, and it follows that 
the GW strain outputs of the two detectors, $h_1(t)$ and $h_2(t)$, 
as given by Eq.~\ref{h_org}, are identical.  The null stream $N_2(t)$ in 
this case is therefore particularly simple:
\begin{equation}\label{eqn:null_h1h2}
N_2(t)=h_1(t)-h_2(t).
\end{equation}
Any GW signal will cancel in $N_2$, up to calibration errors, while 
independent noise in the detectors will not. In this simple case it is 
not even necessary to assume a direction for the incoming GW.

Since the detectors share the same vacuum system,  some noise sources may be correlated in these two interferometers.   An important
motivation for making H2 half the length of H1 in the LIGO design \cite{ligo92} 
was to discriminate against highly correlated disturbances, those that might 
move the mirrors of both interferometers by similar amounts. Such a disturbance
would produce a signal strain twice as large in H2 as in H1, and this would 
indicate that the disturbance was not a gravitational wave. Thus, the 
null-stream veto was built into LIGO from the start.
 
Any larger network containing H1 and H2 could use $N_2$ as a veto. We 
will explore vetos in more general networks, however, since H1 and H2 alone 
do not return polarization or direction information about the source.

\subsection{Three independent detectors at different sites}
We now discuss how to construct null stream using data from three detectors at different sites (e.g., L1, H1, GEO). 
Because of the curvature of the Earth, such detectors 
cannot have exactly coincident antenna patterns.  
The time series of the GW strain plus noise observed
by detectors 1--3 
can be written as a set of three linear equations (the same as 
Equations~4.1a,b,c of \cite{Guersel}),
\begin{eqnarray}
h_1(t)=f^{+}_1(t)h_{+}(t)+f^{\times}_1(t)h_{\times}(t)+n_1(t)\label{h1}
\\
h_2(t_2)=f^{+}_2(t_2)h_{+}(t)+f^{\times}_2(t_2)h_{\times}(t)+n_2(t_2)\label{h2}\\
h_3(t_3)=f^{+}_3(t_3)h_{+}(t)+f^{\times}_3(t_3)h_{\times}(t)+n_3(t_3)\label{h3}
,
\end{eqnarray}
where we define
\begin{equation}\label{eqn:tau}
t_2 = t+\tau_{12}, \qquad t_3 = t+\tau_{13},
\end{equation}
and where in turn $\tau_{1i}$ is the geometrical wave arrival time delay 
expected in detector $i$ with respect to detector 1. We have shown explicitly
the dependence of the antenna pattern functions $f^{+,\times}_i$ on the 
measurement time of the wave at the moving antenna, but we have not 
shown the implicit source-direction 
angular dependence in $\tau_{12}$, $\tau_{13}$, $h_i$, and $f^{+,\times}_i$ 
in order to keep the expressions concise. The intrinsic wave amplitudes 
$h_+$ and $h_\times$ are all evaluated at the same time argument $t$, without 
time delays, because we want to compare the excitation of the various 
detectors by the same part of the incoming wave form; strictly one should 
regard the time-argument of $h_{+,\times}$ as the retarded time along the 
wave's direction of travel. We also include explicitly here 
the noise $n_i$ at the appropriate measurement time in detector $i$. 
In practice, we require that the noise be
stationary and Gaussian within a time scale much longer than the duration of
the triggered events (typically tens of milliseconds for burst GW sources).

It is clear just from counting variables that these three equations admit 
a linear combination that cancels out the GW amplitudes $h_+$ and $h_\times$ 
completely (Equation~4.2 of \cite{Guersel}).  
The null stream for three detectors is then 
\begin{equation}
N_3(\alpha, \delta, t)=A_{23}h_1(t)+A_{31}h_2(t+\tau_{12})+A_{12}h_3(t+\tau_{13}),
\end{equation}
\label{A_null}
where 
\begin{equation}\label{eqn:aij}
A_{ij}(\alpha,\delta,t)=f^{+}_{i}(t_i,\alpha,\delta)f^{\times}_{j}(t_j,\alpha,\delta)-f^{+}_{j}(t_j,\alpha,\delta)f^{\times}_{i}(t_i,\alpha,\delta),
\end{equation}
and where we adopt the convention that $t_1 = t$ the measurement time at our 
reference detector 1. We see that $A_{ij}$ 
is a function of the source direction ($\alpha$, $\delta$) and of time (due 
to the motion of the detectors), 
but it is independent of the polarization angle $\psi$.  Note that, using 
the variables introduced in Equation~\ref{eqn:antpatt}, an alternative and 
somewhat simpler expression for $A_{ij}$ 
\begin{equation}\label{eqn:aijalt}
A_{ij}(\alpha,\delta,t)=A_iA_j\sin(\xi_i-\xi_j),
\end{equation}
where of course $A_i$ and $\xi_i$ depend on $\alpha$, $\delta$, and $t$.

Notice that if two of the three detectors, say 1 and 2, are co-located and 
perfectly aligned, as is the case for H1 and H2, then the coefficient 
$A_{12}$ will vanish while $A_{23}=-A_{31}$. This means that $N_3$ becomes simply proportional to $N_2$, and the three-detector case degenerates to our 
previous two-detector example. More generally, if the source direction is 
such that one of the $A_{ij}$ coefficients is small, which means that the 
corresponding detectors have nearly the same response to the wave, then the 
expression automatically reduces the contribution of the third detector to 
the veto.

\subsection{Two detectors with nearly-aligned antenna patterns}
\begin{figure}
\centerline{\includegraphics[keepaspectratio=true,height=5.4in,angle=0]{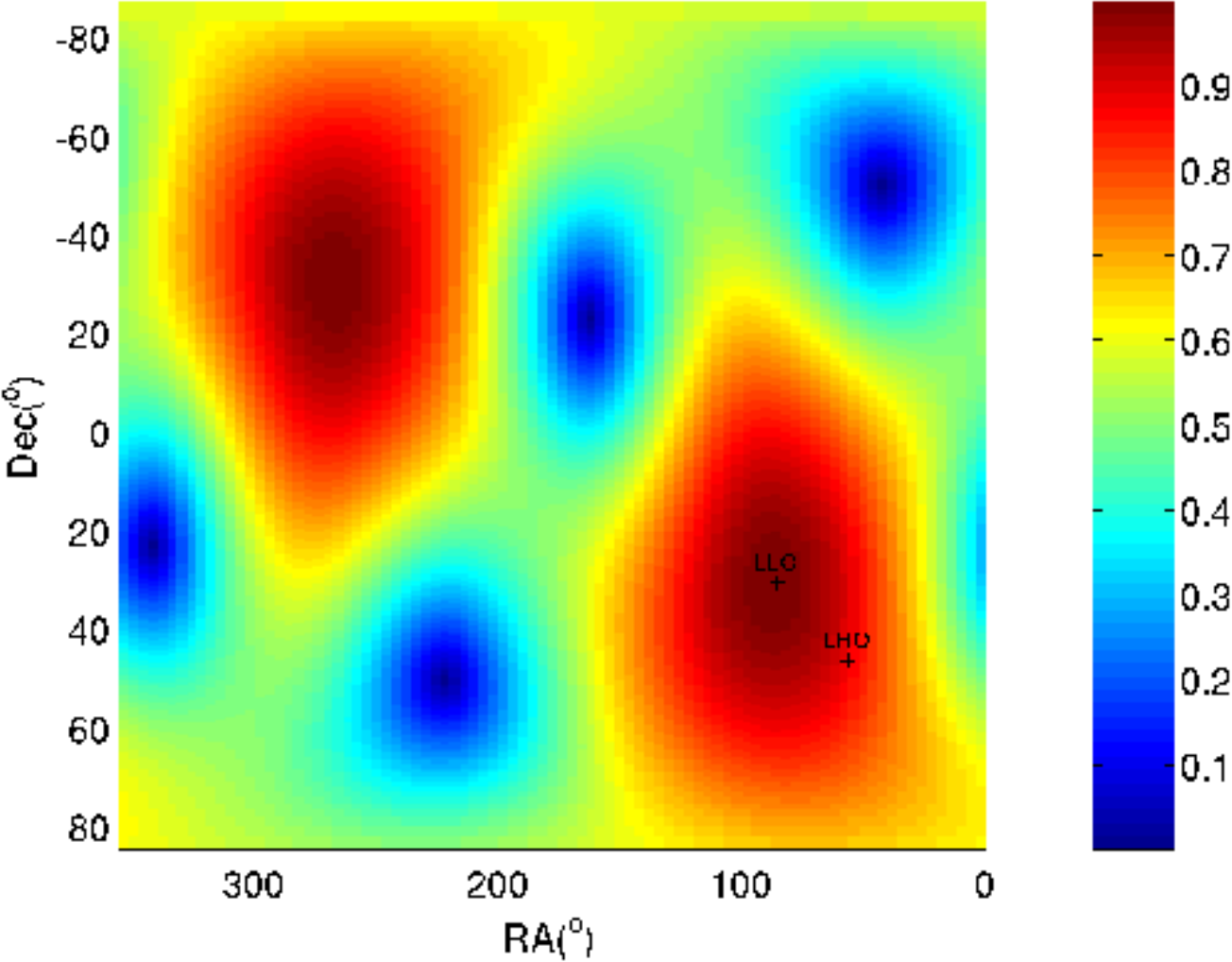}}
\caption{The all-sky map of the amplitude of the antenna beam pattern
  ($A_1$) of LIGO Livingston (L1) at an arbitrary time of 0.00hr, Mar. 18, 2004. The locations for L1
  and H1 are labeled as ``LLO'' and ``LHO'' and marked with symbols ``+''.}
\label{antenna}
\end{figure}

Among currently operating GW detectors, the antenna patterns of L1
and H1 have been designed to be aligned as closely as possible 
with each other, given the curvature of the Earth between them. The sky
directions of their maximum sensitivity ($A_i$) are offset by $\sim
25$ degree (Fig.\ref{antenna}). The two detectors are not exactly 
redundant, as are H1 and H2, but the concept must hold in some 
approximation. We consider here the possibility of constructing a 
``nearly null'' data stream for two nearly perfectly aligned detectors
(i.e., where $A_{12} \sim 0$).  

The residual SNR remaining in any constructed ``nearly'' null stream
unavoidably depends on signal waveforms.  Therefore, the
``best'' null-stream construction method is also waveform-dependent in
general.   In this paper, we consider specifically constructing the
nearly null stream based on a simple
linear combination of data streams from L1 and H1,
\begin{equation}
Q(\alpha, \delta, t) = A_2h_1(t)+\nu A_1h_2(t+\tau_{12}),
\end{equation}
where $\nu$ is a constant depending on source direction only.  We
choose $\nu$ by minimizing the waveform-independent coefficient in the
rms  amplitude of the residual signal 
$\sqrt{\sum_t Q^2(\alpha,\beta,t)}$ over a duration
of time $T$. This yields, in the notation of Equation~\ref{eqn:antpatt}, 
\begin{equation}
\nu=-\cos(\xi_1-\xi_2),
\end{equation}
and the rms amplitude of the residual signal in the null stream is
\begin{equation}
|\sin(\xi_1-\xi_2)|A_1A_2 \sqrt{\sum_t (h^2_{+}(t)+h^2_{\times}(t))\cos^2 (2\psi+\xi^h(t)+\xi_2)}.
\end{equation}
That is, the rms amplitude of the residual signal in the null stream
is proportional to $A_{12}$.  Assuming $A_2>A_1$ and that noise
level of the two detectors are identical, the residual SNR is
approximately  a fraction $f_s\times f_h$ of the original SNR of 
the more sensitive detector, where 
\begin{equation}
f_h=\sqrt{\frac{\sum_t (h^2_{+}(t)+h^2_{\times}(t))\cos^2 (2\psi+\xi^h(t)+\xi_2)}{\sum_{t} (h^2_{+}(t)+h^2_{\times}(t))\sin^2 (2\psi+\xi^h(t)+\xi_2)}},
\end{equation}
and 
\begin{equation}
f_s = \frac{|\sin (\xi_1-\xi_2)|}{\sqrt{A^2_2/A^2_1+\cos^2(\xi_1-\xi_2)}}.
\end{equation}
The quantity $f_h$ depends on characteristics of the wave. 
If the wave is such that 
$\sqrt{h^2_{+}(t)+h^2_{\times}(t)}$ changes more slowly with
time than the phase $\xi^h(t)$, then the average value of $f_h$ will 
be approximately 1. This will be the case for, e.g., waves of 
circular polarization.  

For two perfectly aligned detectors, null stream occurs at
$h_1(t)=-h_2(t+\tau_{12})$, i.e., $\nu=1$ due to the
anti-phase arrangement of one of the arms of L1 and H1.  The null
stream can be constructed on a band of the sky direction 
corresponding to a constant (and
correct) geometrical time delay.  The solution for the sky direction
for two nearly perfectly aligned detectors  degenerates into
(part of) a  ring in the sky corresponding to a constant geometrical
arrival time
delay of the GW at the two detectors.  If the source is near the 
null of the detectors' antenna patterns then the minimum can give an 
incorrect time delay.

\begin{figure}
\centerline{\includegraphics[keepaspectratio=true,height=5.4in,angle=0]{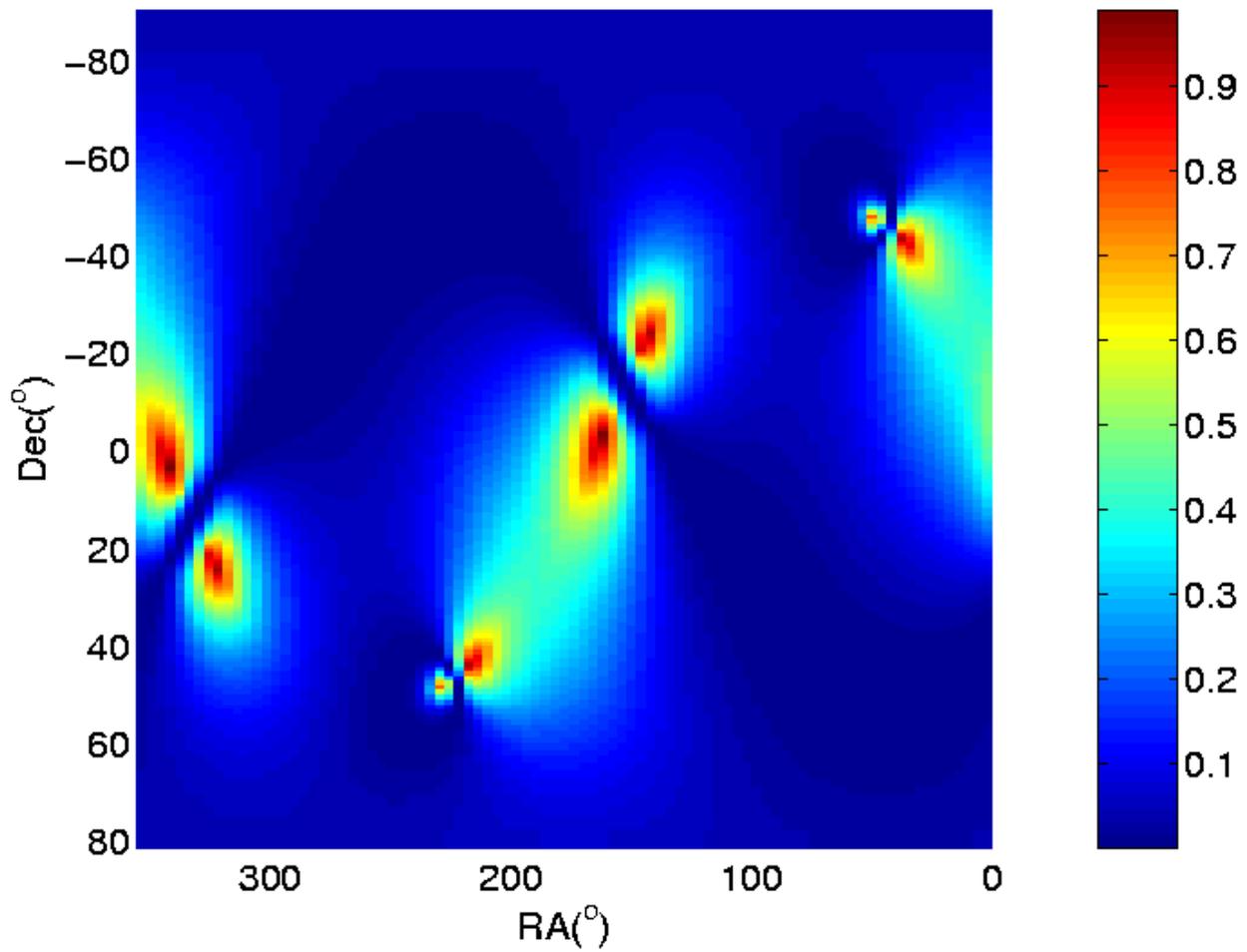}}
\caption{The all-sky map of the waveform-independent fractional SNR
  reduction ($f_s$) in the null stream of the L1-H1 network. The
  fraction is calculated with respect to the SNR of the more sensitive
  detector.}
\label{2_det_fs_all_sky}
\end{figure}

\begin{figure}
\centerline{\includegraphics[keepaspectratio=true,height=5.4in,angle=0]{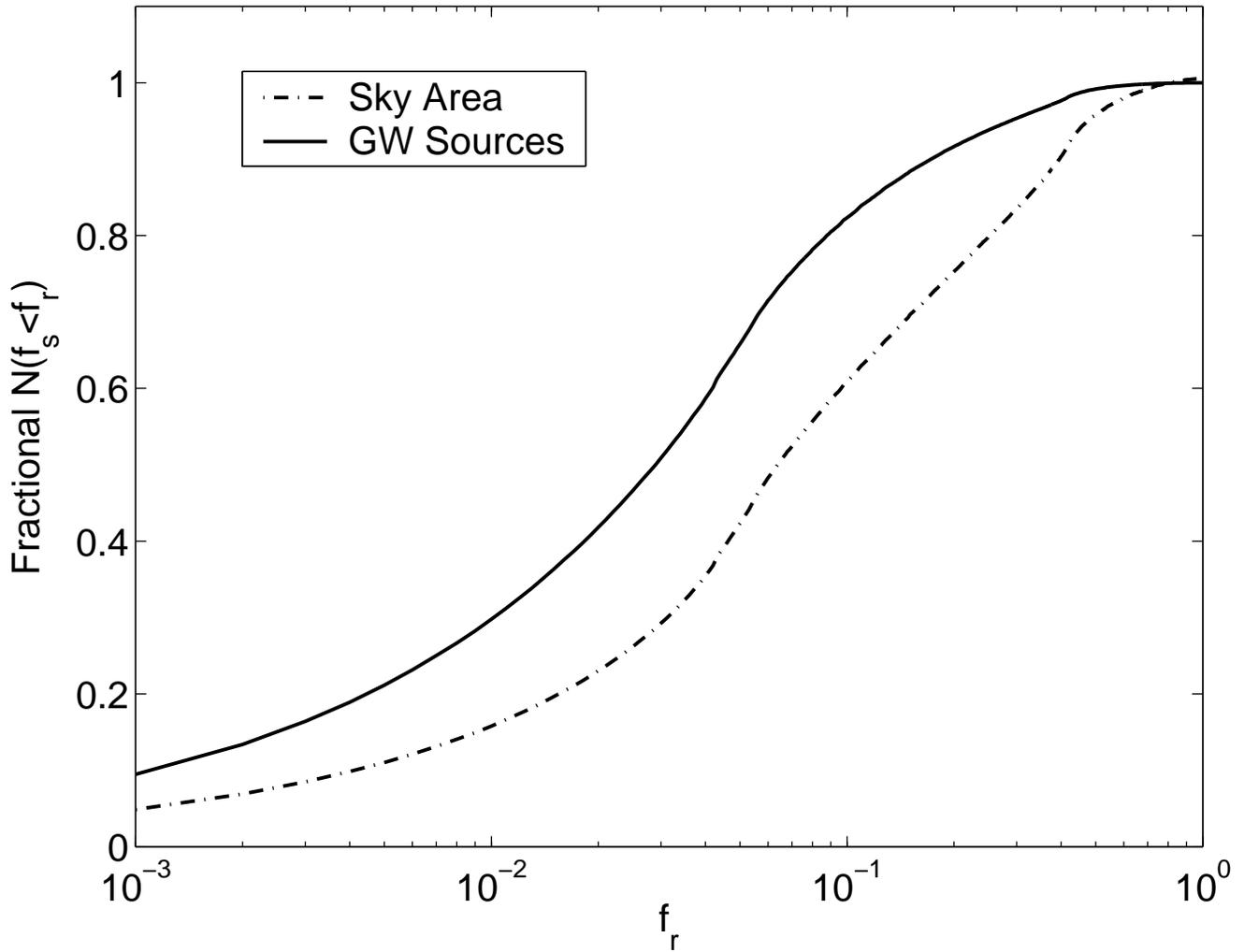}}
\caption{Number distributions in fractions of sky area and number of
  detectable sources as a function of the waveform-independent
  fractional SNR reduction ($f_s$) in the null stream of the L1-H1 network. }
\label{2_det_fs}
\end{figure}

The quantity $f_s$ is independent of wave properties and 
depends only on the sky locations. We show in 
Fig~\ref{2_det_fs_all_sky} an all-sky
map of the value of $f_s$ at an arbitrary time.  Comparing this with 
Fig~\ref{antenna} and Fig.~\ref{2_det_fs}, it is apparent that the
best reduction of SNR occurs around the maximum sensitivities of the
two detector. This is expected, as $A_{12}$ can be best approximated
  as zero in that region.   Figure~\ref{2_det_fs} shows quantitatively 
the sky area
($\int d\alpha d\sin \delta$) vs the fractional reduction ($f_s$).  In
the same figure is also plotted the corresponding fraction of the number
of sources ($\propto \int A^3_I (\alpha,\delta) d\alpha d\sin \delta$),
assuming uniformly distributed sources in flat space-time ($A_I
\propto 1/r$, where the index $I$ refers to the more sensitive
detector, $r$ is the distance to the source). It shows that,  at $f_h \sim 1$, 80\% of sources
(corresponding to 60\% of the sky) can have the SNR reduced by more
than a factor of ten ($f_s<0.1$) in the constructed null stream, and
that 60\% of the sources can have the SNR reduced by more than a
factor of 25 ($f_s <0.04$).   Therefore, we can use our construction of 
the approximate 
null stream to first find sky direction with minimum fractional SNR residual, which can then be checked against the SNR residuals
expected at that sky direction and checked as well for consistency with $f_h$.

Note, however, that this may not be the optimal way of inferring sky
directions. The SNR reductions do indicate the effectiveness of this 
approximate null stream as a veto. 

\section{Implementation}
\label{tech}
Once null data streams are constructed for each of the possible source
directions for an event that passes the coincidence triggers in a burst 
search, we determine the sky directions that show a minimum presence of
the signal, or alternatively the maximum probability that the null stream is
consistent with the expected noise.  If a filter of some kind has been 
applied to the data in order to generate the trigger, then for consistency
one should presumably use this same filter on the null stream before 
making a decision about the presence or absence of a signal.

In this paper,  we adopted a simpler test for the purpose of demonstrating 
the method.  For each sky
direction ($\alpha,\delta$), we construct the following
quantity in
the (discrete) frequency domain for a given duration $T$:
\begin{equation}\label{eqn:power}
P(\alpha, \delta)=2 \sum^N_{k=1}  \frac{|\tilde{N}_k(\alpha,
  \delta)|^2}{\sigma^2_k},
\end{equation}
where $\tilde{N}_k (\alpha, \delta)$ is the Fourier transform of the null
stream $N_3(\alpha,\delta, t)$, $N_2(t)$ or $Q(\alpha, \delta, t)$ over the duration of time $T$,
$\sigma^2_k$ is the expected variance of a true null stream at each
frequency characterized by the noise spectral density $S(f_k)$, $k$ is the frequency index,
and $N <$  half of the number of data points.  To improve the SNR, it is advantageous to
have a band-limited summation in Eqn.~\ref{eqn:power}.  In this paper, we sum over data points
with frequencies $<= 2$ kHz.  In a real search, the frequency
bandwidth can be estimated based on initial triggers from individual detectors.  For the simple H1-H2 null stream, 
$\tilde{N}_k(\alpha,\delta)$ is
replaced with $\tilde{N}_k$ and searches over sky directions or time
delays are not needed. 

For stationary
Gaussian noise, the variance of the noise in the null stream  
for two detectors at the same site is 
\begin{equation}
\sigma^2_k=\sigma^2_{1k}+\sigma^2_{2k}.
\end{equation}
For three independent detectors, the variance is 
\begin{equation}
\sigma^2_k=A^2_{23}\sigma^2_{1k}+A^2_{31}\sigma^2_{2k}+A^2_{12}\sigma^2_{3k},
\end{equation}
while for two nearly aligned detectors at different sites we have 
\begin{equation}
\sigma^2_k=A^2_{2}\sigma^2_{1k}+A^2_{1}\cos^2(\xi_1-\xi_2)\sigma^2_{2k}.
\end{equation}

It follows that, for stationary Gaussian noise, $P(\alpha, \delta)$ is
a random variable drawn from a $\chi^2$ distribution with $2N$ degree of
freedom. In practice, the noise distribution can be determined from the average
properties of data away from the segments we are interested in. (We 
adopt here a strictly frequentist approach. We leave to a later 
investigation the reformulation of this test in Bayesian terms.) We
then search through sky directions (except for the H1-H2 case) for the 
value of
$P(\alpha, \delta)$ that yields the maximum probability to be consistent
with the expected noise distribution.

\section{An example application for BH-BH Merger signals}
\label{example}
\subsection{Simulation elements}
We choose for this simulation a signal with the two-black-hole 
merger waveform obtained from the Lazarus numerical
relativity simulation \cite{baker02}. The
wave is from the merger phase of two 10\msun\ black holes, viewed along
the axis of 
the binary's orbital momentum.  The duration of the wave is about 7 ms
with a central frequency of about 500 Hz. We demonstrate results from two
sky directions, one near L1's maximum sensitivity and the other near its
minimum sensitivity, where GEO600 is as sensitive  as LIGO
detectors at high frequencies despite its smaller size. 

We have adopted the projected detector noise spectral densities $S(f)$ for
initial LIGOs \cite{ligo_SRD},   and for GEO at \cite{geo_Sf} with 500
Hz tuning. Independent Gaussian noise samples at different times were generated 
according to these distributions.  We obtained different
SNRs by varying the distance to the source. The location
information of the different GW observatories were obtained from
\cite{allen96} and references therein.

\subsection{Results}
\begin{figure}
\centerline{\includegraphics[keepaspectratio=true,height=5.4in,angle=0]{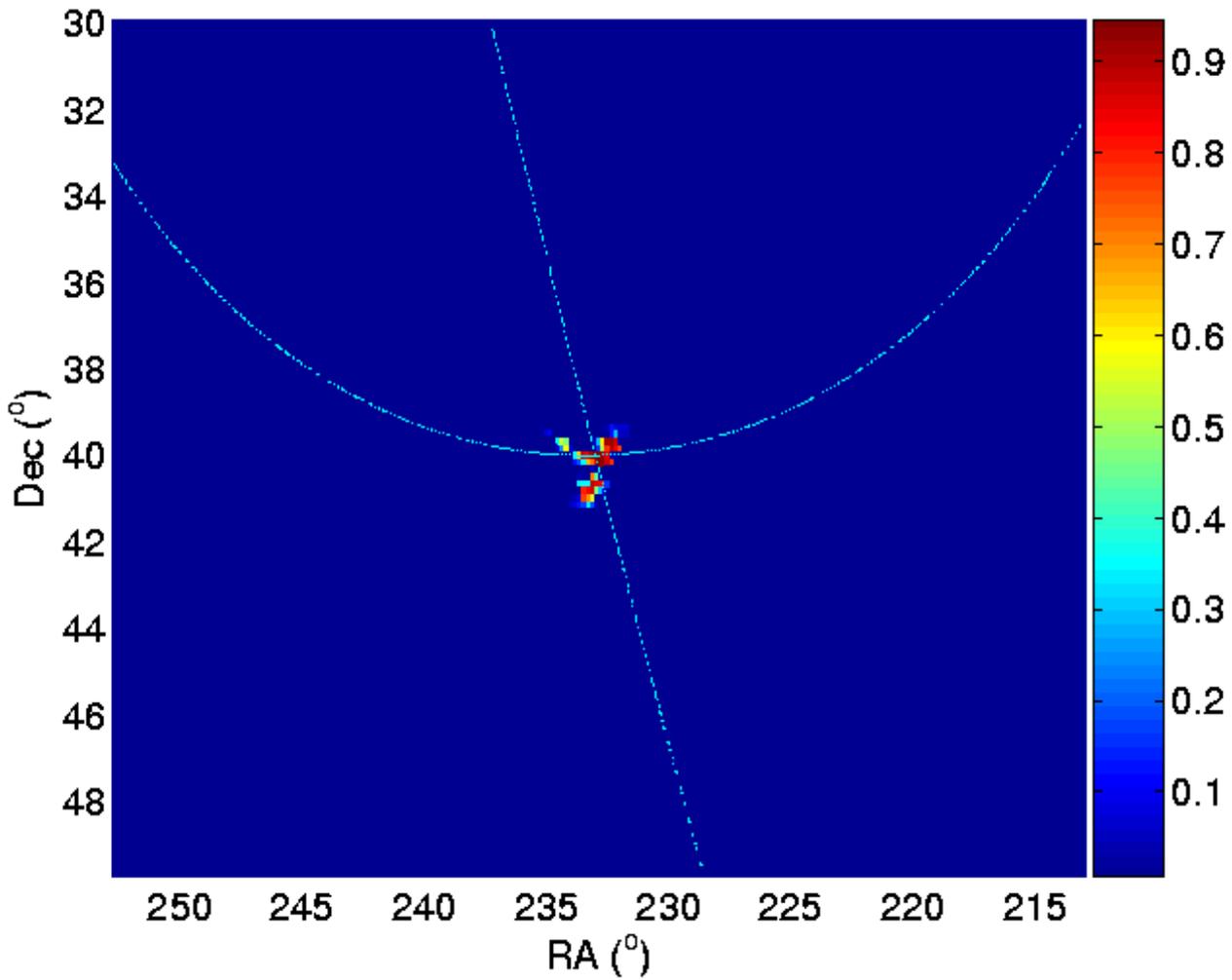}}
\caption{The sky map of the probability that the null stream statistic
  $P(\alpha, \delta)$ constructed for the 3-detector network L1-H1-GEO
  follows a $\chi^2_{2N}$ distribution (see text).  Dotted lines
  indicate the correct time delay contours of the detector pairs L1-H1
  and L1-GEO. The BH-BH merger source is placed at a distance r=1 Mpc
  and in a direction near the null of the L1 sensitivity (center of the plot).  The optimal 3-detector SNR=85.}
\label{3det_min_1M}
\end{figure}

\begin{figure}
\centerline{\includegraphics[keepaspectratio=true,height=5.4in,angle=0]{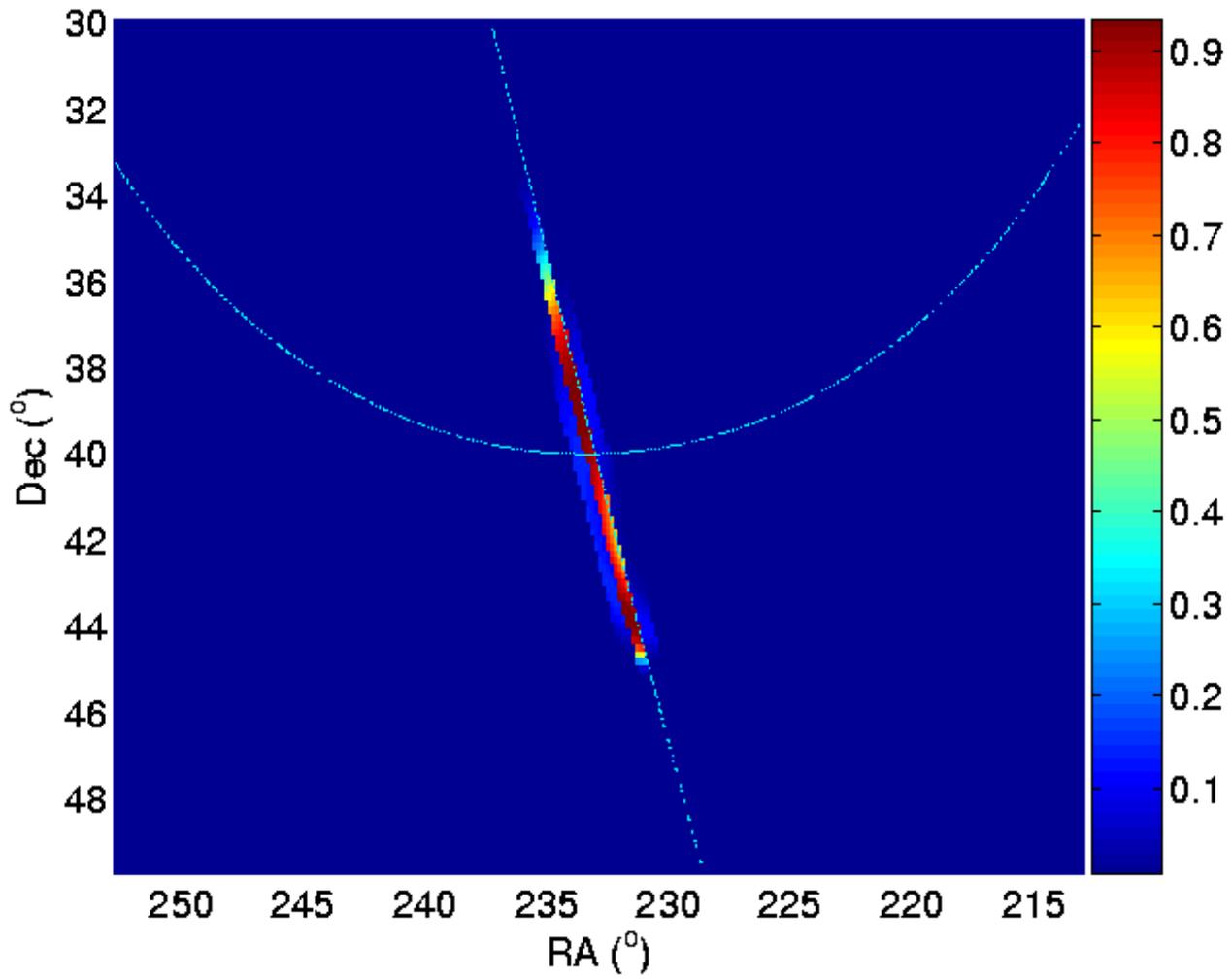}}
\caption{Same as in Fig.~\ref{3det_min_1M}, but for the
  null-stream statistic constructed using the 2-detector network L1-H1, the
  corresponding optimal 2-detector SNR=53.}
\label{2det_min_1M}
\end{figure}

\begin{figure}      
\centerline{\includegraphics[keepaspectratio=true,height=5.4in,angle=0]{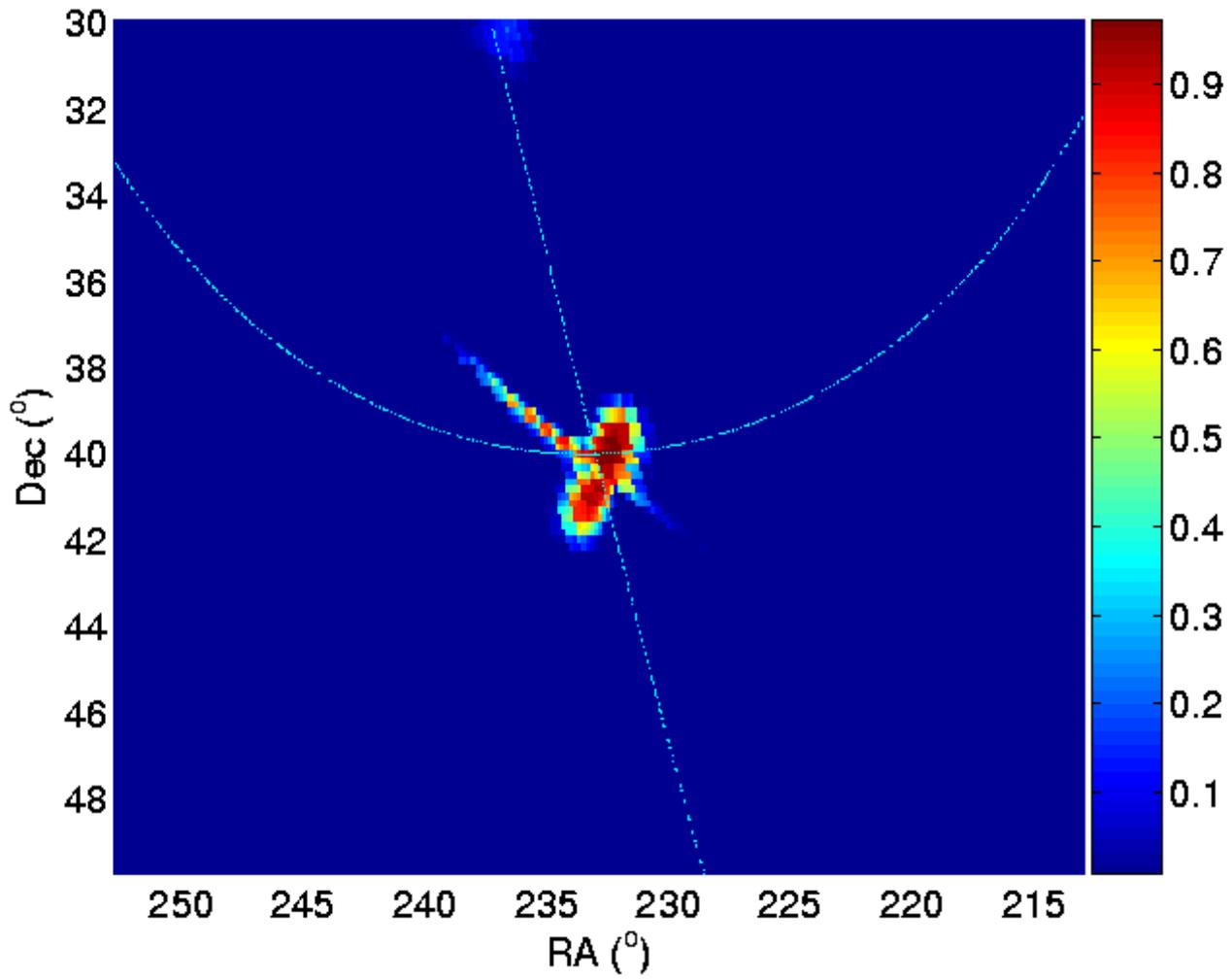}}
\caption{Same as in Fig.~\ref{3det_min_1M} for the L1-H1-GEO network,
  but with the source placed at r=3 Mpc.  The corresponding
  optimal 3-detector SNR=28.5.}
\label{3det_min_3M}
\end{figure}

\begin{figure}
\centerline{\includegraphics[keepaspectratio=true,height=5.4in,angle=0]{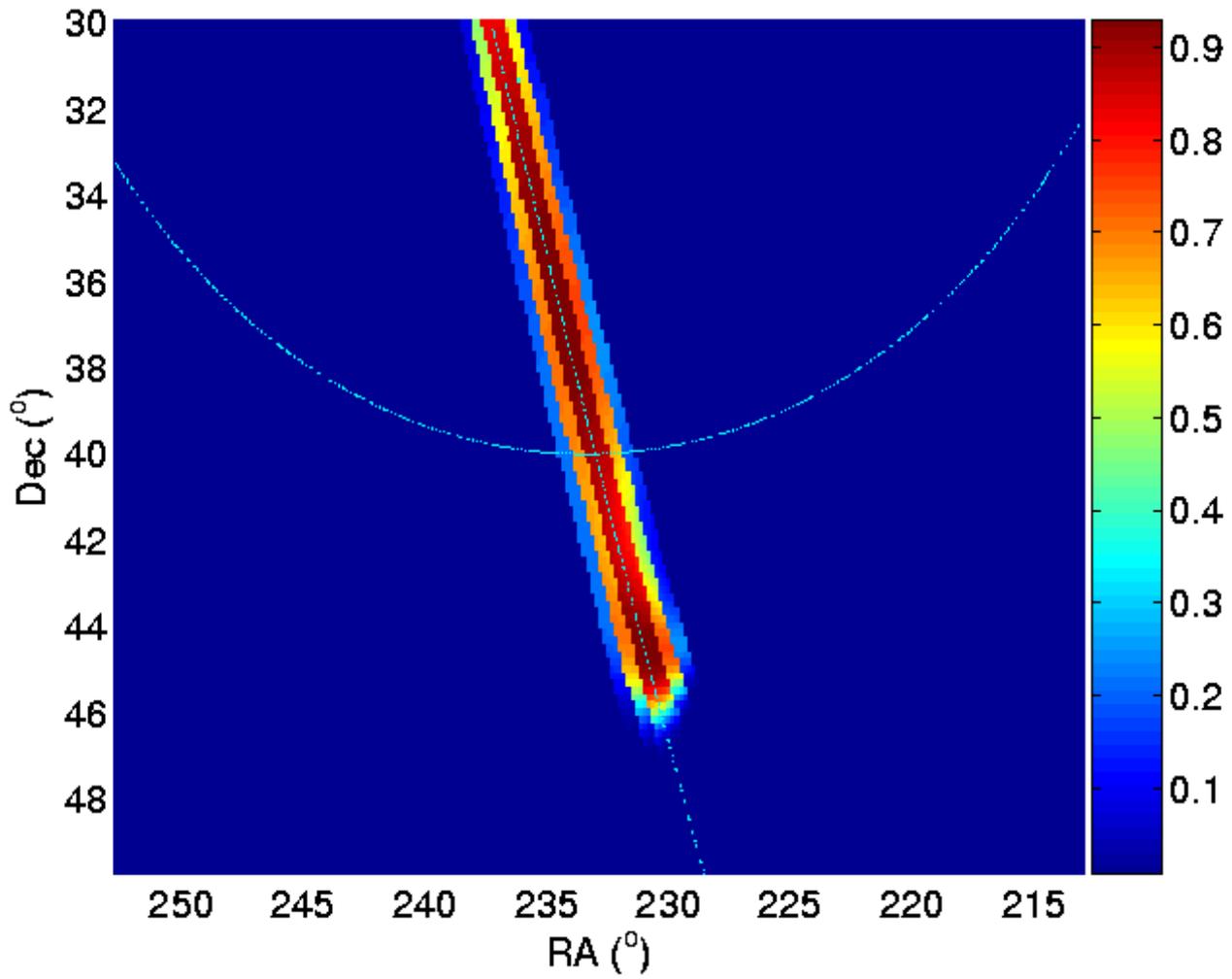}}
\caption{Same as in Fig.~\ref{3det_min_3M} but for the 2-detector
  network L1-H1.  The corresponding optimal two-detector SNR=18.}
\label{2det_min_3M}
\end{figure}

Fig.~\ref{3det_min_1M} and \ref{3det_min_3M} show the source localization using
L1, H1, and GEO. The GW source is placed  at distances of 
1 and 3 Mpc, at a direction near the
minimum L1 sensitivity. The arrival time of the GW at L1 was chosen
arbitrarily to be at 0.00 hr,  March, 18, 2004.  The optimal SNR quoted here is defined as
the optimal combination of the signal-to-noise ratios $\rho_i$ from the matched
filtering technique $\rho=\sqrt{\sum_i \rho^2_i}$.  In these two figures,
$\rho=85$ ($\rho_1=47$, $\rho_2=23$, $\rho_3=67$) and 28.5. Source direction can be determined from
fractions of a degree to a few degrees.  
Figs.~\ref{2det_min_1M} and \ref{2det_min_3M}
show the results from the same data but using the two-detector network
of L1 and H1 only.  The optimal SNR is 53 and 18. The source direction 
can be determined in one dimension from fractions of a 
degree to several degrees. 

\begin{figure}
\centerline{\includegraphics[keepaspectratio=true,height=5.4in,angle=0]{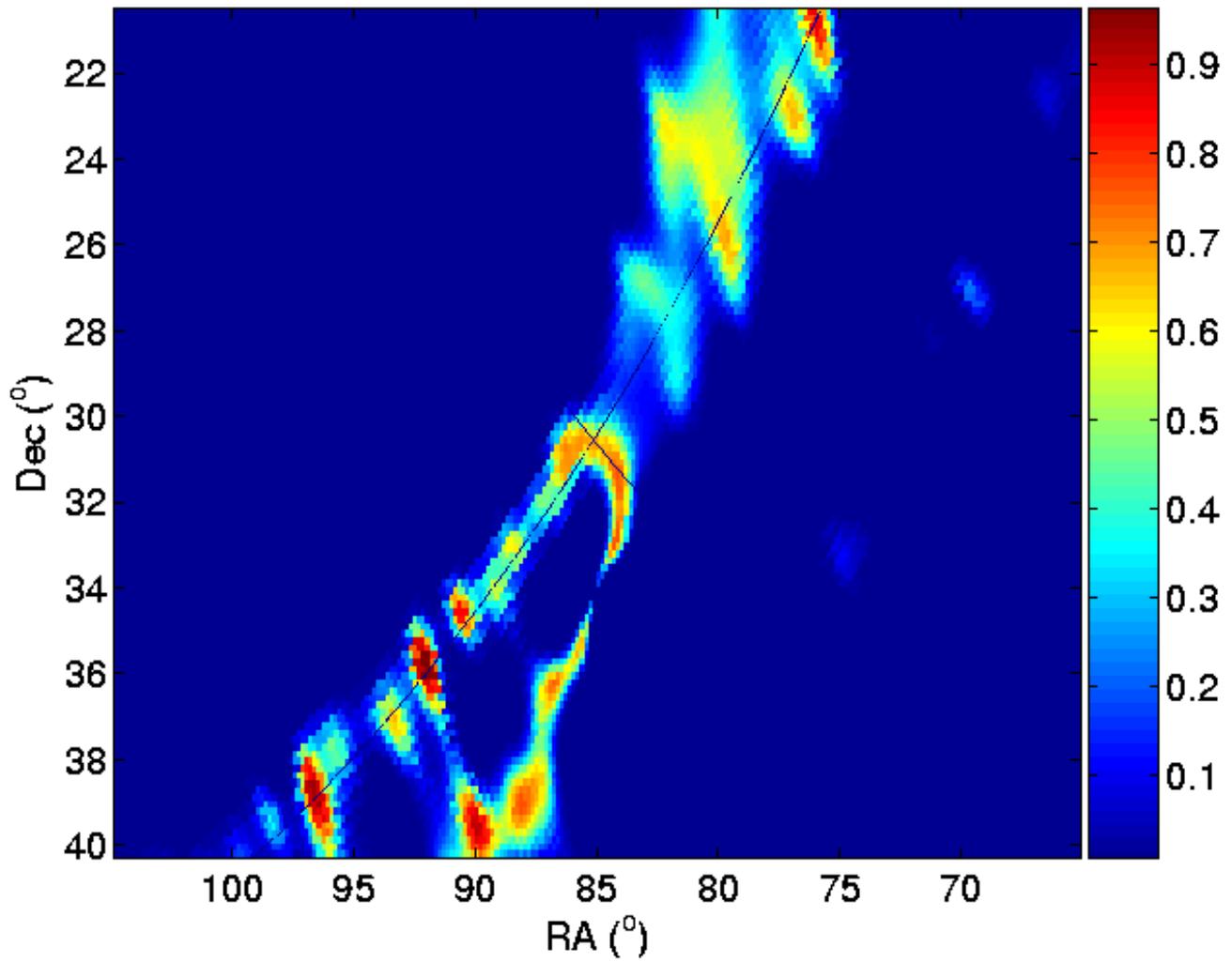}}
\caption{Same as in Fig.~\ref{3det_min_1M} for the 3-detector network
  L1-H1-GEO, but with source placed near the maximum L1
  sensitivity (center of the plot), source distance r=10 Mpc, and the corresponding optimal
  3-detector SNR=20, SNR in GEO is 2.}
\label{3det_max_10M}
\end{figure}

\begin{figure}
\centerline{\includegraphics[keepaspectratio=true,height=5.4in,angle=0]{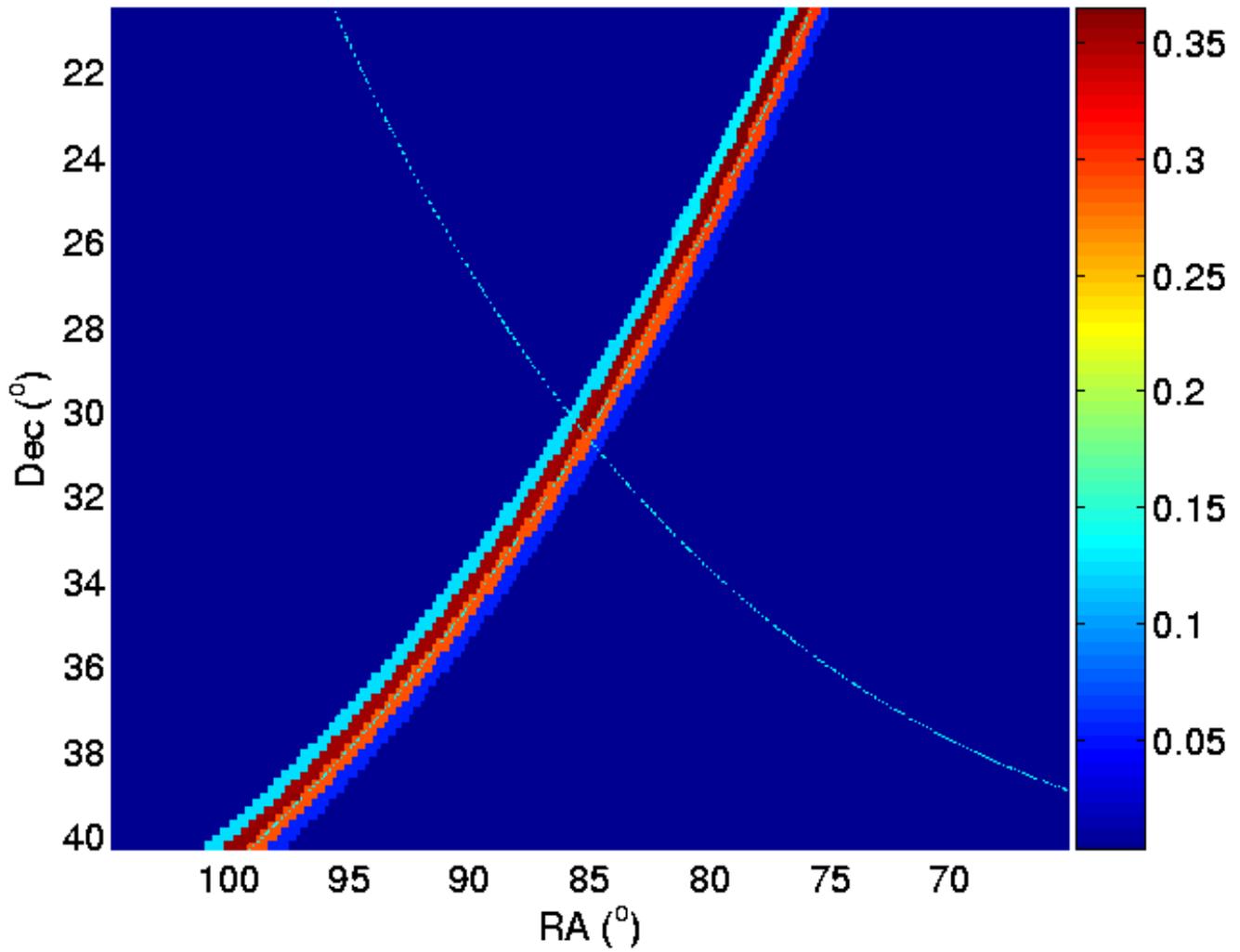}}
\caption{Same data as in Fig.~\ref{3det_max_10M}, but for the
  2-detector network L1-H1. The corresponding optimal 2-detector SNR=20.}
\label{2det_max_10M}
\end{figure}

\begin{figure}
\centerline{\includegraphics[keepaspectratio=true,height=5.4in,angle=0]{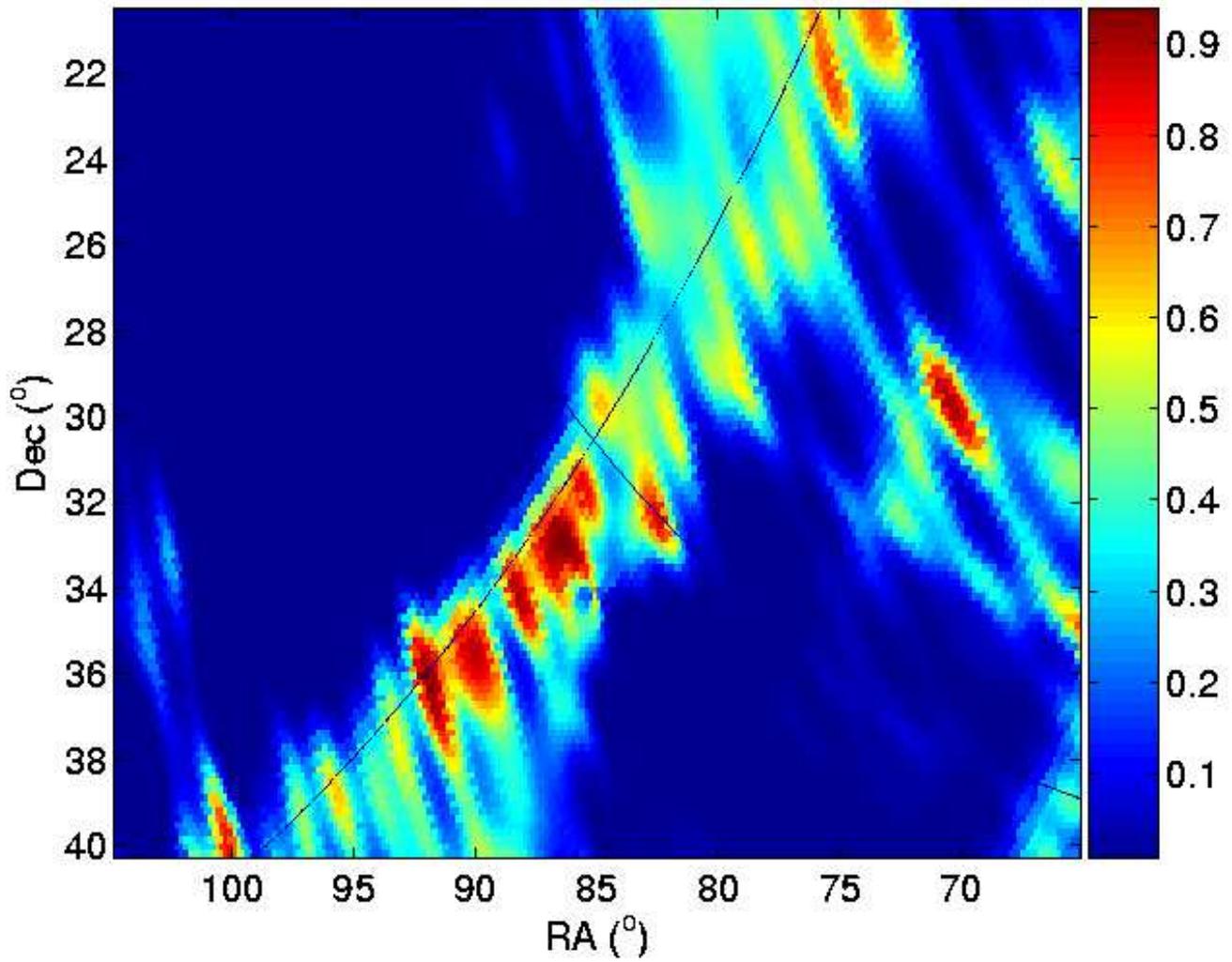}}
\caption{ Same as in  Fig.~\ref{3det_max_10M} for the L1-H1-GEO
  detector network but with the source distance r=20 Mpc and
  SNR=10. SNR in GEO is 1.}
\label{3det_max_20M}
\end{figure}

\begin{figure}
\centerline{\includegraphics[keepaspectratio=true,height=5.4in,angle=0]{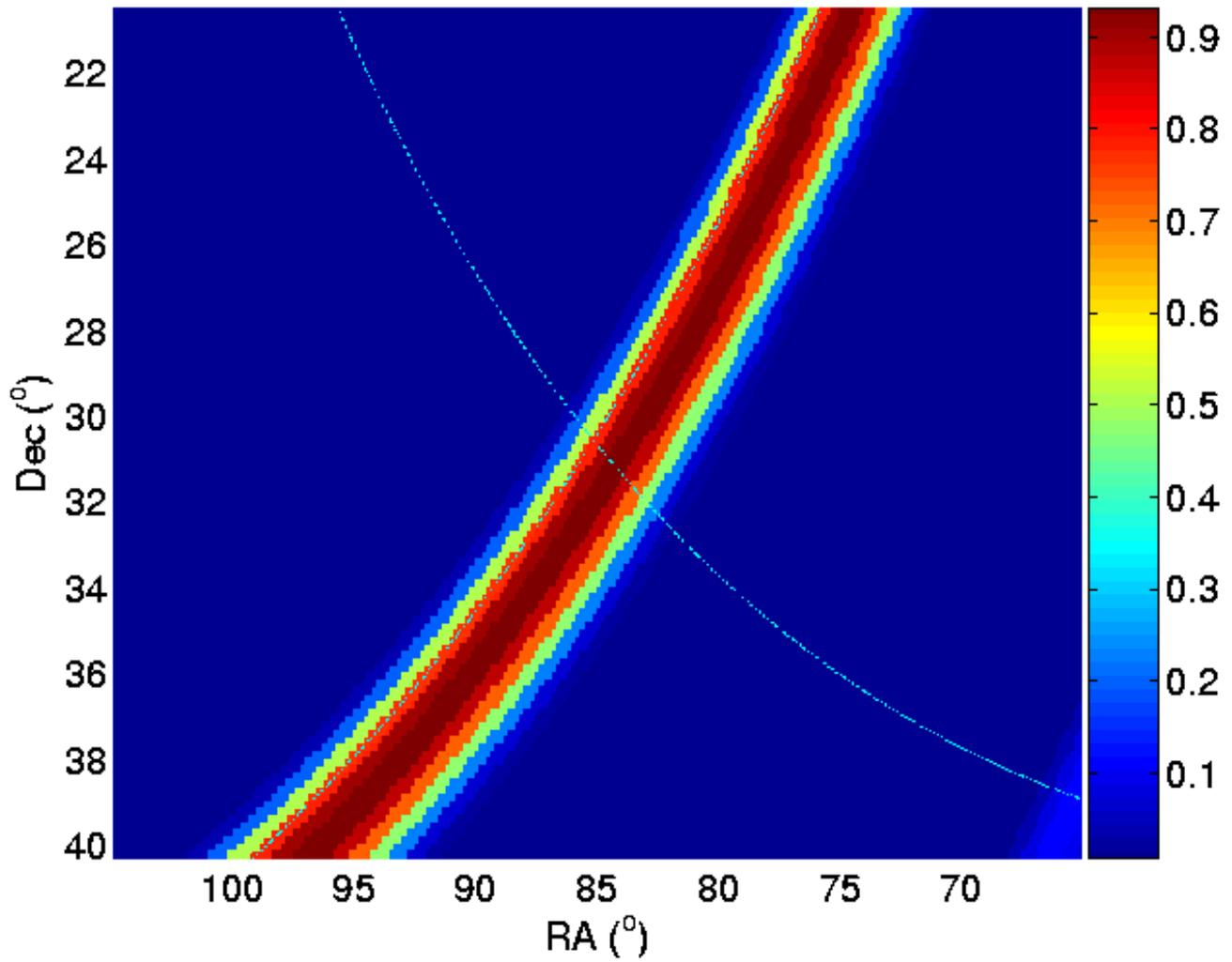}}
\caption{The same data as in Fig.~\ref{3det_max_20M} but for the
 2-detector network L1-H1, source distance r=20 Mpc and SNR=10.}
\label{2det_max_20M}
\end{figure}

Fig.~\ref{3det_max_10M} and \ref{3det_max_20M} show the source localization
using L1, H1, and GEO but with source placed  at distances of 
10, 20 Mpc, at the direction
near  the maximum L1 sensitivity. The optimal 
SNR $\rho$ is 20 ($\rho_1=15$, $\rho_2=13$, $\rho_3=2$) and 10. In these two cases, the SNR from GEO is
around 2 and 1. That is, GEO is not very sensitive to the signals.  As
a result, the spatial resolution is very poor for a
3-detector network.  Figs.~\ref{2det_max_10M} and \ref{2det_max_20M}
show the results of the same data using L1 and H1 only (optimal SNRs
are 20 and 10). The sky direction can be determined to degrees along
one dimension.

Note that in  the three-detector network, when the SNR is high enough, the
sky location can be uniquely determined. This is an improvement from
the degeneracy in localization with time delay information only.  
 
\section{Conclusion}
\label{discussion}

We have presented the principle of constructing null streams using data
 from two or three detectors that are redundant or almost so. 
The null streams, which contain no GW signal, are useful as vetos and 
as localizers of source directions, as demonstrated by the examples.  The source localizations can be a useful 
consistency check on the veto, in the sense that an event that passes the 
veto should also localize as expected; failure to do so might indicate other
problems in the data analysis.  In turn, the veto is likely to 
be very effective at discriminating non-Gaussian noise events from real ones. 
The simplest veto to use is the comparison of the two LIGO Hanford detectors, 
but other vetos could be used for various configurations, such as the two
large LIGO detectors along with GEO600. When only the two large LIGO 
detectors are being used, there still seems to be an approximate null stream 
that is effective over most of the sky.

Notice that the null stream construction method can also be used to test the
validity of the theory of General Relativity (GR).  General relativity predicts that there are only two
transverse wave polarizations in GWs ($h_+$, and
$h_\times$ in this context).  As a result, a true null stream can be
constructed for any waveforms 
using three detectors at different sites and the two wave
polarizations can be extracted for sufficient strong signals.   On the
other hand, scalar-tensor theory 
gravitational waves can contain a transverse breathing
mode in addition to these two transverse modes \cite{Will}.  
In this case, a null
stream can be constructed for any waves with at least four detectors 
at different
sites and the breathing mode, if present with sufficient
strength, can be measured using three detectors. If the 
four-detector null stream shows no signal but null streams constructed
from subsets of three detectors do, then this would indicate that a scalar 
polarization mode is present in the wave. Even if only three detectors 
are available, it might be possible to detect a scalar mode from the 
failure of the null stream, provided that the GW event can be confirmed
by other means.

This paper is only a preliminary investigation of what we believe will become a powerful tool for data analysis. Many issues remain
to be studied. We plan in subsequent publications to address: 
(1) use of the veto from H1 and H2 in realistic data analysis, 
    taking into account calibration error, spatial and timing resolution, 
    and the effect of nonlinear trigger algorithms;
(2) understanding under what circumstances the H1-L1-GEO veto will 
    be useful and whether the null stream will help localize, and the 
    effect of including the VIRGO detector in three- and four-detector null 
    streams, and investigating practical issues such as the best use 
    of information on time delays and signal durations;
(3) consideration of incorporating the null-stream search algorithm
    into an all-sky coherent detection method by adding 
    data together to maximize SNRs at the same sky
    location where null streams are constructed (one method already proposed
    in \cite{Guersel, anderson01}); 
(4) signal recovery and parameter estimation; 
(5) computational cost and search grid size;  and 
(6) application to testing alternative theories for gravitational waves.

\ack
We would like to thank Peter Saulson for carefully reading through our
manuscript and for his critical comments.  We also thank B. S. Sathyaprakash
for very useful and inspiring discussions on this work.

\clearpage
\newpage


\end{document}